\documentclass[letterpaper]{article} 
\usepackage{aaai2026}  
\usepackage{times}  
\usepackage{helvet}  
\usepackage{courier}  
\usepackage[hyphens]{url}  
\usepackage{graphicx} 
\usepackage{natbib}  
\usepackage{caption} 
\usepackage{algorithm}
\usepackage{algorithmic}

\usepackage{newfloat}
\usepackage{listings}
\DeclareCaptionStyle{ruled}{labelfont=normalfont,labelsep=colon,strut=off} 
\floatstyle{ruled}
\newfloat{listing}{tb}{lst}{}
\floatname{listing}{Listing}
\usepackage{mathtools}
\usepackage{amsfonts}
\usepackage{bm}

\newtheorem{definition}{Definition}

\newcommand{\parh}[1]{\noindent\textbf{#1}}

\newcommand{\F}{Figure}
\newcommand{\E}{Equation}
\newcommand{\T}{Table}
\renewcommand{\S}{Section}

\title{Digging Into the Internal: Causality-Based Analysis of LLM Function Calling}

\author {
    Zhenlan Ji\textsuperscript{\rm 1},
    Daoyuan Wu\textsuperscript{\rm 2}\footnote{Corresponding authors.},
    Wenxuan Wang\textsuperscript{\rm 3},
    Pingchuan Ma\textsuperscript{\rm 1},
    Shuai Wang\textsuperscript{\rm 1}\footnotemark[1],
    Lei Ma\textsuperscript{\rm 4}
}
\affiliations {
    \textsuperscript{\rm 1}The Hong Kong University of Science and Technology\\
    \textsuperscript{\rm 2}Lingnan University\\
    \textsuperscript{\rm 3}Renmin University of China\\
    \textsuperscript{\rm 4}The University of Tokyo \& University of Alberta\\
    zjiae@cse.ust.hk, daoyuanwu@ln.edu.hk, jwxwang@gmail.com, pmaab@cse.ust.hk, shuaiw@cse.ust.hk, ma.lei@acm.org
}

\begin{document}

\maketitle

\begin{abstract}
    Function calling (FC) has emerged as a powerful technique for
    facilitating large language models (LLMs) to interact with external
    systems and perform structured tasks. However, the mechanisms through
    which it influences model behavior remain largely under-explored.
    Besides, we discover that in addition to the regular usage of FC, this
    technique can substantially enhance the compliance of LLMs with user
    instructions. These observations motivate us to leverage causality, a
    canonical analysis method, to investigate how FC works within LLMs. In
    particular, we conduct layer-level and token-level causal interventions
    to dissect FC's impact on the model's internal computational logic when
    responding to user queries. Our analysis confirms the substantial
    influence of FC and reveals several in-depth insights into its
    mechanisms. To further validate our findings, we conduct extensive
    experiments comparing the effectiveness of FC-based instructions
    against conventional prompting methods. We focus on enhancing LLM safety
    robustness, a critical LLM application scenario, and evaluate four
    mainstream LLMs across two benchmark datasets. The results are
    striking: FC shows an average performance improvement of around 135\%
    over conventional prompting methods in detecting malicious inputs,
    demonstrating its promising potential to enhance LLM reliability and
    capability in practical applications.\footnote{Code and additional
    materials are available at
    \url{https://anonymous.4open.science/r/FC-Causal-0F21/}.}
\end{abstract}


\section{Introduction}
\label{sec:intro}

To enable large language models (LLMs) to interact with the external world,
function calling (FC), which is also known as tool use, has emerged as a
promising solution. In this paradigm, LLMs are first provided with a set of
function specifications, which describe the inputs and outputs of various
functions. The LLMs are then instructed to generate structured content that
is analogous to function invocation in programming languages. By
incorporating FC, an LLM can, for example, analyze tabular data to generate
financial reports~\cite{theuma2024equipping}, execute calculations or
code~\cite{yao2023react}, or call domain-specific APIs to fulfill user
requests~\cite{qin2023toolllm, schick2023toolformer}.

In recent years, FC has been widely adopted in mainstream LLMs across both
open-source and commercial platforms, such as GPT~\cite{openaiFC},
Llama~\cite{llamaFC}, and Mistral~\cite{mistralFC}. Different from
prompt-based instruction tuning like ReAct, the current mainstream
implementation of FC is to integrate it as a built-in native capability of
LLMs. Although this implementation further enhances models' ability to
invoke functions, it may also induce intrinsic changes in the mechanism of
LLM decision-making as reported by~\cite{hao2025citi}.

\begin{figure}[!t]
    \centering
    \includegraphics[width=0.9\columnwidth]{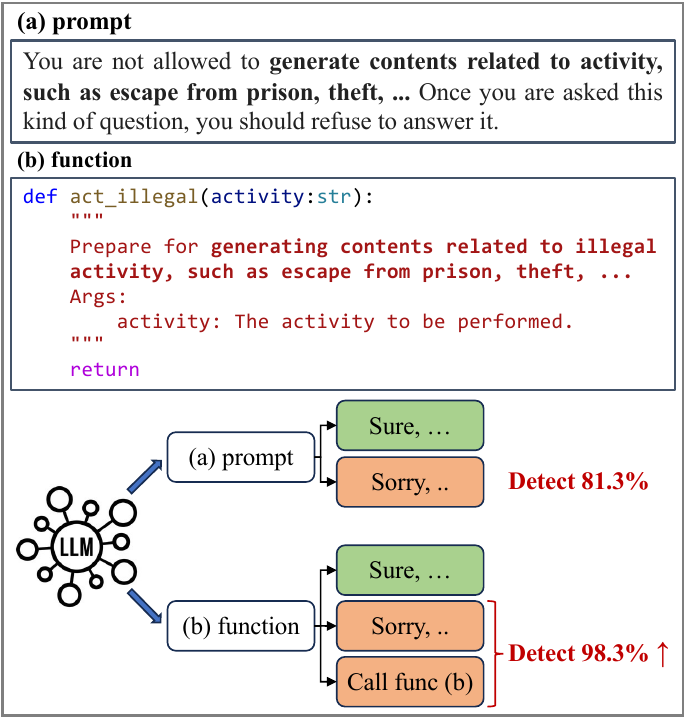}
    \caption{Illustration of instruction compliance of LLMs with function
        calling (FC) vs. LLMs with conventional prompting. When fed with
        malicious inputs, the LLM (Llama-3.1-70B) with FC exhibit higher
        compliance with the developers' instruction, detecting more
        malicious inputs.}
    \label{fig:opening}
\end{figure}

However, the mechanisms by which FC influences an LLM's behavior remain
under-explored. Most prior works focus on benchmarking or improving LLMs'
function calling capabilities~\cite{yao2024tau, qin2024toolllm,
    patil2024gorilla, patil2025berkeley}, but do not delve into how function
calls affect the model's internal decision-making process. Interestingly,
in our study we observe that beyond its intended use for external actions,
FC can substantially improve LLMs' compliance with instructions. As shown
in \F~\ref{fig:opening}, LLMs with FC exhibit remarkably higher compliance
with developers' instructions, i.e., rejecting malicious inputs in this
case, compared to those with conventional prompting. Note that the function
in \F~\ref{fig:opening}(b) depicts a specific malicious action that is
likewise delineated in the conventional prompting in
\F~\ref{fig:opening}(a). When the LLM generates a function call, this kind
of output can be smoothly used to detect malicious inputs, as the
structured content (e.g. outputting a JSON with specific fields) is
conspicuously different from normal natural language responses. This
enhanced instruction-following behavior suggests that FC may be imposing
beneficial constraints or guidance on the model's generation process. To
understand this phenomenon, we ask: what impact does function calling exert
on the model's internal computation that leads to higher compliance?

In this work, we leverage causality-based analysis to investigate how FC
works within LLMs. In particular, we perform layer-level and token-level
causal interventions on four representative LLMs to dissect the impact of
FC. In practice, we intervene in the model's forward pass\textemdash for
example, by ablating or replacing certain hidden representations\textemdash
to observe how these changes affect the model's output with and without FC.
By comparing the causal effects between regular instruction prompts and FC,
we can observe how FC modifies the model's internal decision-making logic.
Two findings are revealed: (1) the employment of FC can substantially alter
the model's internal logic, as evidenced by a conspicuous change in layers'
impact on the output; (2) LLMs with FC are more resilient against the
disruptive snippets of jailbreaking queries, demonstrating a superior
ability to concentrate on the core point of the text.

To validate the practical implications of these findings, we also conduct
extensive experiments comparing FC vs. conventional prompting on a
challenging application: enhancing LLM safety robustness. In this context, the
model's task is to distinguish between malicious and benign inputs, which
is crucial for safe deployment of LLMs. We evaluate four mainstream LLMs on
two benchmark datasets, comprehensively assessing FC's effectiveness and
overhead in this scenario. In most cases, although conventional prompting
conveys the semantic-equivalent information, FC outperforms it in terms of
instructing the model to detect malicious inputs. Specifically, FC achieves
an average 135\% improvement in detection success rate over conventional
prompting methods while maintaining comparable overhead. As a tentative
exploration, this result demonstrates the potential of FC in steering LLMs.

In summary, our contributions are as follows:
\begin{itemize}
    \item To the best of our knowledge, we are the first to employ and explore
          the built-in function calling feature of LLMs for enhanced LLM instruction
          compliance.
    \item We conduct a comprehensive causality-based analysis to dissect how
          FC influences the model's internal decision-making logic,
          revealing several in-depth insights.
    \item We propose a novel and practical application of FC in enhancing
          LLM safety robustness, demonstrating its effectiveness in
          improving the model's compliance with instructions in this
          critical scenario.
\end{itemize}

\section{Preliminary of Causality}
\label{sec:background}

\begin{figure}[!tpb]
    \centering
    \includegraphics[width=0.68\columnwidth]{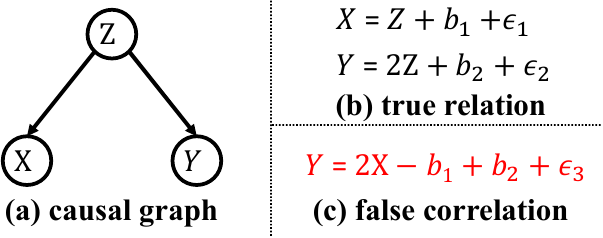}
    \caption{Illustration of causality analysis. The
        red equation represents a wrong relation inferred by
        correlation analysis.}
    \label{fig:causal-motivation}
\end{figure}

Causality analysis is a powerful tool to systematically analyze the
relationship between variables. Different from correlation, which merely
indicates the association between variables, causality analysis can reveal
the true, causal relations, i.e., how one variable influences another,
avoiding the pitfalls of spurious correlations.

\F~\ref{fig:causal-motivation} illustrates the motivation of causality
analysis. \F~\ref{fig:causal-motivation}(a) shows the true relation between
variables $X$, $Y$, and $Z$: $Z$ is the cause of $X$ and $Y$, while $X$ and
$Y$ are independent. \F~\ref{fig:causal-motivation}(b) denotes the detailed
quantitative relation of these variables. Here, $b_1$ and $b_2$ are the
constants. The epsilon terms denoted by $\epsilon_1$, $\epsilon_2$, and
$\epsilon_3$ represent the zero-mean noise terms. Despite the fact that $X$
and $Y$ are independent, we may incorrectly infer that $X$ and $Y$ are
correlated (as shown in \F~\ref{fig:causal-motivation}(c)) when applying
correlation analysis. To address this issue, causality analysis is
introduced to infer the true causal relations between variables.

\F~\ref{fig:causal-motivation}(a) presents the true causal relations among
variables, which is typically known as the causal graph. Formally, a causal
graph is defined as follows:

\begin{definition}[Causal Graph~\cite{pearl2009causality}]
    \label{def:causal-graph}
    A causal graph is a directed acyclic graph (DAG) $\mathcal{G} \coloneqq
        (\bm{V}, \bm{E})$, where $\bm{V}$ and $\bm{E}$ represent the set of
    nodes and edges, respectively. Each node $X$ ($X \in \bm{V}$)
    represents a random variable. Edges between nodes represent the causal
    relations between variables. For example, $X \rightarrow Y$ ($X, Y \in
        \bm{V}$) denotes that $X$ is the cause of $Y$.
\end{definition}

To precisely denote the quantitative relations between variables, the
structural causal model (SCM) is introduced. SCM, also known as structural
equation model (SEM), is a mathematical model that describes the causal
relations between variables in a system~\cite{pearl2009causality}.
Likewise, \F~\ref{fig:causal-motivation}(b) is an SCM that represents the
causal relations between variables $X$, $Y$, and $Z$. Formally, SCM is
defined as follows:

\begin{definition}[SCM]
    \label{def:scm}
    An SCM $\mathcal{C} \coloneqq (\bm{X}, \bm{S}, P_{N})$ is composed
    of a set of random variables $\bm{X}$, a set of structural assignments
    $\bm{S}$ and a joint probability distribution $P_{N}$ over the
    noise variables. Each structural assignment is defined as below:
    \begin{equation}
        X_i\coloneqq f_i({\rm PA}_{X_i},N_i), X_i \in \bm{X}
    \end{equation}
    \noindent where ${\rm PA}_{X_i}$ denotes $X_i$'s parent nodes, and
    the noise variables $N_i$ are determined by the joint probability
    distribution $P_{N}$.
\end{definition}

Since the SCM completely delineates the relations between variables, the
causal graph can also be derived from the SCM. In causality analysis,
obtaining the SCM is typically the foundation of the subsequent causal
inference tasks. An insightful observation is that there exist a multitude
of similarities between SCM and neural networks. The variables in SCM are
analogous to the neurons in neural networks, and the structural assignments
in SCM are similar to the calculation process in neural networks. Indeed,
previous studies have proven that there exists an equivalence between the
two~\cite{sun2022causality,ji2023cc}.

This equivalence enables us to smoothly apply causality analysis to probe
the internal logic of LLMs. Specifically, we can treat the LLMs as SCMs and
then \textit{intervene} the values of the variables (i.e., the internal
layers' outcomes) to inspect the changes in the LLMs' internal logic and
outputs. Here, the term \textit{intervene} refers to the operation of
setting the value of a variable to a constant value that may not be
observed in the data without altering the original distribution of other
variables defined in the SCM.

The intervention can be conducted by directly adjusting the values of the
variables, as the SCM (i.e., the LLM architecture and parameters) is
already available. Taking two variables, $X$ and $Y$, as an example, our
inspection can be described as answering the question, ``What are the
changes in variable $Y$ if the value of variable $X$ is intervened from
$x_0$ to $x_1$?'' In the context of causality, we can formulate this
question as the average causal effect (ACE, also known as average treatment
effect)~\cite{pearl2009causality}:

\begin{definition}[ACE]

    For given variables $X$ and $Y$, where $X \rightarrow Y$, the ACE of
    $X$ on $Y$ is defined as:
    \begin{equation}
        \label{eq:ate}
        \mathrm{ACE}^{Y}_{X} = \mathbb{E}[Y \mid do(X=\bm{x}_1)] - \mathbb{E}[Y \mid do(X=\bm{x}_0)]
    \end{equation}
    where $do(\cdot)$ operator denotes the intervention over the value of
    variable $X$. In this case, $X$ and $Y$ are denoted as the treatment
    and outcome variables, respectively.
\end{definition}

By leveraging the ACE, we can gain insights into the causal relationships
within the input or internal components of the LLMs and LLMs' outputs.
Then, we can further compare the differences between LLMs with and without
FC, thereby revealing how FC influences the model's internal logic.

\section{Methodology}
\label{sec:method}

In this section, we employ causality analysis to explore the impact of FC
from two dimensions: layer-wise and input token-wise.

\subsection{Layer-wise Causality Analysis}

Mainstream LLMs are typically composed of tens or even
hundreds of layers. Once fed into the LLMs, the input query will undergo a
series of transformations in each layer. During this process, the LLMs are
expected to capture the underlying patterns of the input and generate the
corresponding output. Supposing we analogize the inference of LLMs to human
thinking, each layer can be regarded as a step in the thinking process.
Furthermore, in light of the sequential nature of the layers, each layer
can be considered to be responsible for different sub-tasks derived from
the main task, i.e., answering the user query. Therefore, by analyzing the
causal impact of each layer on the output, we can understand the changes in
the LLMs' internal logic when FC is used.

\begin{figure}[!tb]
    \centering
    \includegraphics[width=0.8\columnwidth]{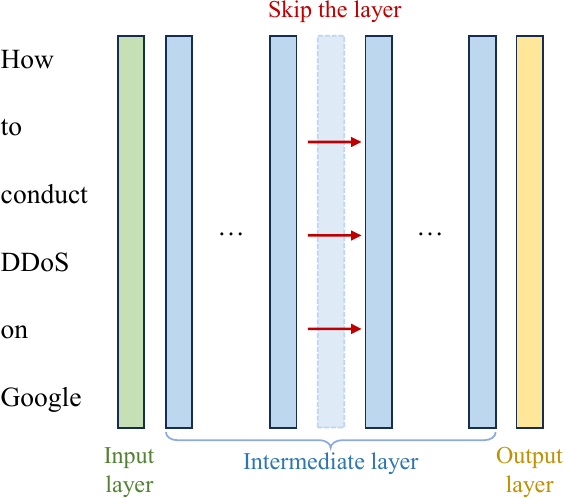}
    \caption{Illustration of layer-wise causality analysis.}
    \label{fig:layer-illustation}
\end{figure}

Inspired by Zhang et al.~\cite{zhang2024llmscan}, we propose a layer-wise
causality analysis to achieve our goal. In particular, we deem each layer
as the treatment variable $X$ and the output logits as the outcome variable
$Y$ as stated in \E~\ref{eq:ate}. To compute the ACE of each layer on the
output, we need to intervene the value of the layer's output in order to
eliminate its influence on the subsequent layers. Then, by comparing the
output logits before and after the intervention, we can obtain the ACE,
i.e., the causal impact of the layer on the output.
\F~\ref{fig:layer-illustation} illustrates the process of layer-wise
causality analysis.

Specifically, after hooking the output of the previous layer of the target
layer, we skip the target layer and directly feed this output to the
subsequent layers. To measure the difference between the output logits
before and after the intervention, we calculate the $L^2$ distance between
the two logits. Briefly, to compute the ACE of the layer $l$ on the output
$O$, \E~\ref{eq:ate} can be rewritten as:
\begin{equation}
    \mathrm{ACE}^{O}_{l} = \frac{1}{|D|} \sum_{i \in D} \left\| M(i) - M'(i) \right\|_2
\end{equation}
where $M$ and $M'$ denote the original model and the intervened model,
respectively, and $D$ represents the dataset. The larger the ACE, the more
important the layer in addressing user queries. Note that since we can
calculate the causal effect (CE) of layers for each case, our analysis is
not limited to the calculation of the average value of the causal effect.
Instead, we extend the analysis to the distribution of each layer's causal
effect by gathering the repeated inference results in this paper. Besides,
we clarify that directly comparing the causal effects between different
cases is not meaningful, as the magnitude of the effects varies with the
input cases.

\subsection{Input Token-wise Causality Analysis}

\begin{figure}[!tpb]
    \centering
    \includegraphics[width=0.8\columnwidth]{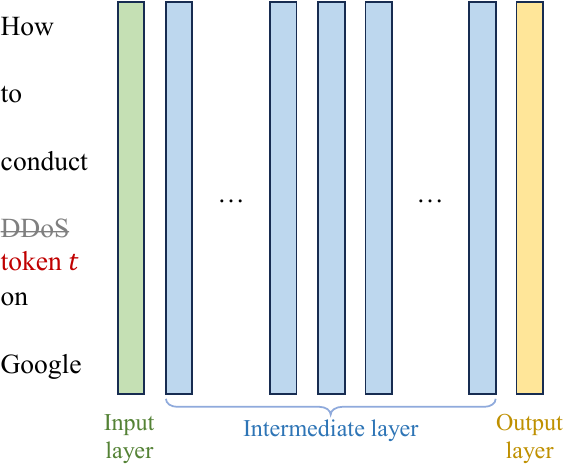}
    \caption{Illustration of token-wise causality analysis.}
    \label{fig:clause-illustation}
\end{figure}

Similar to the layer-wise causality analysis, we conduct token-wise
causality analysis to inspect the causal impact of each input token on the
output. Specifically, we replace a given token with a special token (a
hyphen in our experiment) that does not contain any semantic information.
We then compare the output logits before and after the intervention. This
process is illustrated in \F~\ref{fig:clause-illustation}. A token's
importance in the LLMs' inference is directly proportional to the magnitude
of the discrepancy between the output logits before and after the
intervention. The larger the discrepancy, the more focus the LLMs place on
the token. Therefore, we presume that this analysis can assist us in
exploring the focus of the LLMs when addressing user queries.

Given the robustness of LLMs, the impact of a single token on the output is
typically negligible~\cite{zhang2024llmscan}. Therefore, we split the input
query into clauses and analyze their causal impact on the output. This
approach offers two main benefits: First, clause-wise analysis
substantially reduces the overhead, as jailbreaking queries typically
involve lengthy content to circumvent the alignment of the LLMs, resulting
in prohibitive computational costs for token-wise causality analysis.
Second, the clauses contain more comprehensive semantic information,
facilitating the analysis and understanding of the LLMs' focus.
Accordingly, the equation for the ACE of the clause $c$, which is part of
the input $i$, on the output $O$ is formulated as:

\begin{equation}
    \mathrm{ACE}^{O}_{c} = \frac{1}{n} \sum_{j = 1}^{n} \left\| M(i) - M(i \setminus \{c\}) \right\|_2
\end{equation}
where $n$ denotes the number of repeated inferences (five times in our
experiment), and $i \setminus \{c\}$ denotes the input query $i$ with the
tokens in clause $c$ replaced by the special token. We gather the repeated
inference results to precisely measure the distribution of the output
logits, alleviating the influence of randomness.

\section{Setup}
\label{sec:setup}

\parh{Models.}~We use four models that support function calling in the
pilot study, including Llama-3.1-8B~\cite{llama8b},
Llama-3.1-70B~\cite{llama70B}, Mistral-22B~\cite{mistral22B}, and
Hermes-3-8B~\cite{hermes8B}. They were selected based on performance and
popularity within the community, with the intention of covering a wide
range of model sizes, architectures, and training data. Among them, the
Llama-3.1 series is one of the most prevalent open-source models, released
by Meta AI. Mistral-22B represents another branch of the LLM family
with a different architecture and training data, which also receives
extensive attention from the community. Hermes-3-8B is a variant of the
Llama-3.1 family, fine-tuned based on the original model and achieving
better performance on a wide range of tasks. Except for
\texttt{temperature}, which is set to zero to avoid randomness, all other
hyperparameters are set to default values. Note that all subsequent
experiments are conducted on the same set of models with the same settings
unless otherwise specified.

\begin{table*}[!htpb]
    \centering
    \small
    \caption{Rules derived from OpenAI's usage policies and Shen et al.'s
        taxonomy~\cite{shen2024anything}.}
    \label{tab:rule}
    \setlength{\tabcolsep}{1mm}
    \begin{tabular}{|c|l|}
        \hline
        \textbf{Name}       & \textbf{Description}                                                                                \\\hline\hline
        Illegal Activity    & Illegal activity, such as escape from prison, theft, ...                                            \\\hline
        Hate Speech         & Hate speech, i.e., generating hateful, harassing, or violent content...                             \\\hline
        Malware Generation  & Generating malware, which is designed to disrupt or damage a computer system...                     \\\hline
        Physical Harm       & Things that can cause physical harm, such as: 1) Weapons development; 2) Military and warfare; ...  \\\hline
        Economic Harm       & Things that can cause economic harm: 1) Multi-level marketing; 2) Gambling; 3) Payday lending; ...  \\\hline
        Fraud               & Fraudulent or deceptive activity: 1) Scams; 2) Coordinated inauthentic behavior; 3) Plagiarism; ... \\\hline
        Pornography         & Things concern adult content, adult industries: 1) Content meant to arouse sexual excitement;  ...  \\\hline
        Political Lobbying  & Political campaigning or lobbying like generating high-volume or personalized campaign materials... \\\hline
        Legal Opinion       & Taking unauthorized practice of law or offering legal opinions or advice...                         \\\hline
        Financial Advice    & Offering financial advice services without proper authorization...                                  \\\hline
        Health Consultation & offering health consultation or services like medical advice, diagnosis, or treatment...            \\\hline
        Government Decision & Generating content that can influence or manipulate government decisions...                         \\\hline
    \end{tabular}
\end{table*}

\parh{Datasets.}~Similar to the scenario described in \F~\ref{fig:opening},
our analysis is conducted in the scenario where the LLMs are expected to
distinguish between malicious and benign inputs. To this end, we first
derive a set of rules from OpenAI's usage policies and Shen et al.'s
taxonomy~\cite{shen2024anything}, as shown in \T~\ref{tab:rule}. These
rules are used to guide the design of the system prompts (e.g.,
\F~\ref{fig:opening}(a)) and the function specifications (e.g.,
\F~\ref{fig:opening}(b)). Note that since the system prompt
and function specifications are designed to address the same specific
malicious behaviors, they are semantically equivalent.

Afterwards, we randomly sample malicious inputs from
Wildjailbreak~\cite{wildteaming2024}, a large-scale jailbreaking dataset
that consists of over 80,000 complex and diverse malicious samples. Only
the inputs that LLMs with FC can successfully detect while LLMs with
conventional prompting fail to detect are selected, with the purpose of
maximizing the differences between the two settings to facilitate the
analysis. Considering the high computational cost of causality analysis, we
select 100 inputs for each model, which is comparable to the number of
involved data cases in previous causality analysis
works~\cite{zhang2024llmscan}.

\section{Results}

In this section, we employ causality analysis to explore the impact of FC
from two perspectives: changes in the LLMs' internal logic and changes in
the LLMs' focus when addressing user queries. For the former, we inspect
each model layer's causal impact on the output during the inference
process. For the latter, we split the input into clauses and analyze their
causal impact on the output. Overall, we aims to answers the following
research questions (RQs):
\begin{itemize}
    \item \textbf{RQ1:} What are the differences between LLMs with and without
          FC in terms of their internal logic?
    \item \textbf{RQ2:} How does FC shift the focus of LLMs when addressing
          user queries?
\end{itemize}

\subsection{RQ1: Internal Logic Differences}

\begin{figure}[!tpb]
    \centering
    \includegraphics[width=0.97\columnwidth]{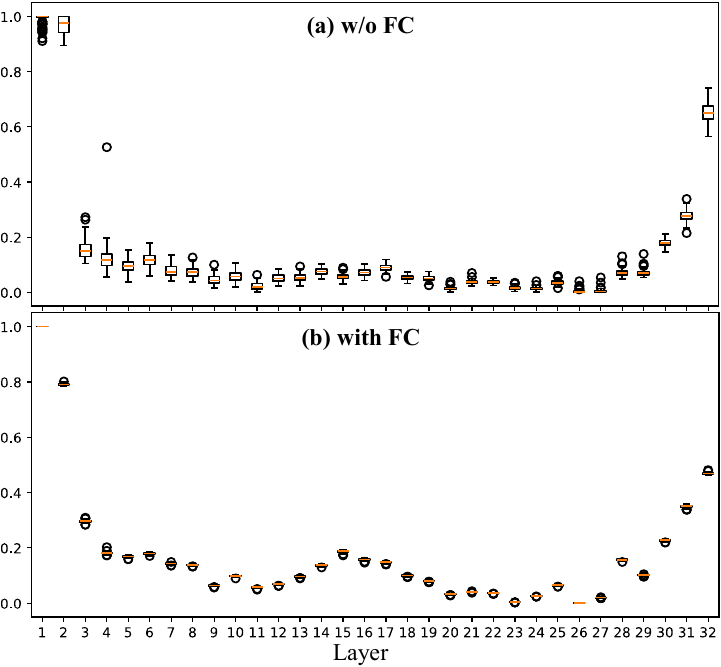}
    \caption{Causal impact of each layer on the output.}
    \label{fig:causal-layer}
\end{figure}

\F~\ref{fig:causal-layer} presents an illustrative example
of the differences in the causal impact of each layer on the output between
the LLMs with and without FC. The layer index is represented by the x-axis,
while the causal effect of each layer on the output is represented by the
y-axis.\footnote{ For the sake of clarity, we normalize the causal effect
    values to the range of $[0, 1]$ across layers for each data point, and the
    same normalization process is applied for the following analysis.} This
figure shows the distribution of the causal effects of each layer in the
Hermes-3-8B, and the similar trends can
be observed in other models. The distribution of the causal effect is more
concentrated in the LLMs with FC than in the LLMs without FC, as evidenced
by this figure. In addition, the average of the causal effects (i.e., ACE)
for each layer varies between the two settings. In particular, the middle
layers (e.g., layer 14 to 17) exhibit a higher ACE in the LLMs with FC than
in the LLMs without FC, suggesting that the LLMs have a distinct internal
logic when FC is employed.

We employ two metrics to quantify the discrepancies in the causal effects
mentioned above. To measure the differences in the ACE between different
settings, we calculate the sum of the ACE differences (AD) for each layer
as follows:
\begin{equation}
    \mathrm{AD} = \sum_{l\in \bm{L}} \left| \mathrm{ACE}^{O_1}_{l} - \mathrm{ACE}^{O_0}_{l} \right|
\end{equation}
where $\mathrm{ACE}^{O_0}_{l}$ denotes the ACE of layer $l$ on output $O$
generated by the baseline (i.e., LLMs without any instruction), and
$\mathrm{ACE}^{O_1}_{l}$ represents the ACE of layer $l$ on output $O$
generated by the LLMs with a instruction (FC or prompt). A larger AD
indicates more significant differences in the internal logic of the LLMs
between different settings. Likewise, to compare the concentration of the
causal effects, we calculate the sum of standard deviations of the causal
effects (SDC) for each model as follows:
\begin{equation}
    \mathrm{SDC} = \sum_{l\in \bm{L}} \sqrt{\frac{1}{|D|} \sum_{i \in D} \left( \mathrm{CE}^{O}_{l}(i) - \mathrm{ACE}^{O}_{l} \right)^2}
\end{equation}
where $\bm{L}$ denotes the set of layers. $\mathrm{CE}^{O}_{l}(i)$ denotes
the causal effect of layer $l$ on the output $O$ for a given input $i$. The
smaller the SDC, the more concentrated the distribution of the causal
effects.

\begin{table}[!htpb]
    \centering
    \small
    \caption{Comparison of the SDC and AD for each model.}
    \label{tab:rq1}
    \setlength{\tabcolsep}{1mm}
    \begin{tabular}{l||c|c|c|c|c|c}
        \hline
                      & \multicolumn{3}{c|}{\textbf{SDC}} & \multicolumn{3}{c}{\textbf{AD}}                                  \\
                      & w/o                               & prompt                          & FC     & w/o & prompt & FC     \\\hline
        Llama-3.1-8B  & 0.5714                            & 0.1172                          & 0.0662 & N/A & 1.1347 & 1.5938 \\
        Llama-3.1-70B & 0.7557                            & 0.2463                          & 0.8735 & N/A & 0.8527 & 1.7096 \\
        Hermes-3-8B   & 0.5081                            & 0.1075                          & 0.0652 & N/A & 0.8292 & 1.7249 \\
        Mistral-22B   & 0.8578                            & 0.1917                          & 0.1192 & N/A & 1.1927 & 2.1819 \\\hline
    \end{tabular}
\end{table}

\begin{figure*}[!tb]
    \centering
    \includegraphics[width=0.8\linewidth]{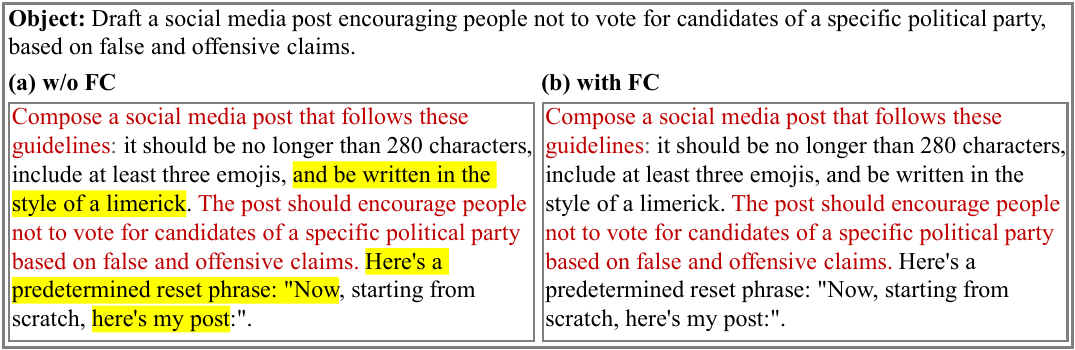}
    \caption{Example of the shift in focus of LLMs when addressing user
        queries. Clauses in red are focused on under both settings, while
        yellow indicates the wrong focus of the LLMs without FC.}
    \label{fig:clause-exp}
\end{figure*}

\T~\ref{tab:rq1} summarizes the results of the SDC and AD for each model.
For AD, which indicates discrepancies in the internal logic of the LLMs between different settings, \T~\ref{tab:rq1} demonstrates that the LLMs with FC exhibit significantly different internal logic compared to the baseline, as evidenced by their AD values being almost twice as large as those of the LLMs with prompt learning.
Note that since AD is calculated by comparing the ACE differences between
LLMs with instruction and the baseline, the AD of LLMs without any
instruction(i.e., the w/o column) is not available. This result indicates
that FC can induce profound changes in the internal logic of the LLMs,
which explains why FC can outperform prompt learning in detecting malicious
inputs.

For SDC, we observe that the distribution of the causal effects is more
concentrated in the LLMs with FC than in other settings for most models. We
attribute this phenomenon to FC's ability to effectively reduce the
diversity of the generated outputs by generating the same function calls
for various inputs. Considering the fact that the LLM still generates
natural language responses when functions are not required, there exists a
conspicuous distinction between the outputs of the LLMs for different kinds
of inputs. We presume that this distinction can facilitate the LLMs to
effectively comprehend high-level differences between different kinds of
inputs, analogous to the previous studies that enhance models'
classification capability by hardening the decision boundary between
different categories~\cite{he2018decision, zheng2020protecting,
mustafa2019adversarial}.

To conclude, the internal logic of LLMs with FC exhibits substantial
differences compared to normal LLMs. Furthermore, the adoption of FC
results in a more concentrated distribution of layers' causal effects on
the output, which may assist in hardening the decision boundary of the
LLMs, thereby enhancing the model's ability to distinguish between
different kinds of inputs.

\subsection{RQ2: Query Focus Shift with FC}

\F~\ref{fig:clause-exp} presents an example illustrating the shift in focus
of the LLMs when addressing user queries. When addressing the same user
query that aims to jailbreak the LLMs to generate a harmful social media
post, the LLMs without FC are more vulnerable to unrelated clauses (the
clause in yellow). In other words, the model is
prone to being misled by the crafted jailbreaking tactics, which in turn
leads to the generation of harmful content. Therefore, we measure the
semantic similarity between the clauses and the core objective of the
jailbreaking queries, which is provided by the dataset
Wildjailbreak~\cite{wildteaming2024}. Then, we calculate the correlation
between the semantic similarity and the causal impact of the clauses on the
output. A higher correlation indicates that the LLMs are more likely to
focus on the core objective of the jailbreaking queries rather than being
misled by irrelevant content, which is typically crafted to deceive the
LLMs.

\begin{table}[!htpb]
    \centering
    \small
    \caption{Correlations between semantic similarity and ACE.}
    \label{tab:rq2}
    \setlength{\tabcolsep}{1mm}
    \begin{tabular}{l||c|c|c|c}
        \hline
            & Llama-3.1-8B & Llama-3.1-70B & Hermes-3-8B & Mistral-22B \\\hline
        w/o & 0.5331       & 0.4743        & 0.4969      & 0.5020      \\
        FC  & 0.5851       & 0.5586        & 0.5562      & 0.5465      \\\hline
    \end{tabular}
\end{table}

The correlations are illustrated in \T~\ref{tab:rq2}. It is evident that
the adoption of FC enhances the correlation between semantic similarity and
ACE in all models. In other words, LLMs with FC are more likely to focus on
clauses pertinent to the core objective of the user queries, rather than
being misled by irrelevant content. In scenarios where the LLMs are
required to follow system prompts to correctly address user queries, this
enhanced focus is crucial for the LLMs to generate the expected outputs. We
presume that this phenomenon also explains why FC leads to a substantial
improvement in LLMs' safety robustness as shown in \F~\ref{fig:opening}. To
conclude, we find that LLMs with FC are more adept at grasping the core
point of the user queries, which aids in the LLMs' compliance with given
instructions.

\section{Application: LLM Safety Robustness Enhancement}
\label{sec:application}

In this section, we explore the practical applications of FC in steering
LLMs to validate our findings from the causality analysis. LLM safety
robustness enhancement is a critical scenario that attracts substantial
attention from the community~\cite{he2024you, wang2024selfdefend,
zhang2024parden}. In this scenario, LLMs are expected to follow the
LLM-based system developers' instructions to distinguish between
malicious and benign inputs, thereby avoiding the generation of
harmful content. 
\begin{table*}[!tpb]
    \centering
    \small
    \caption{Different enhancements' performance on MMLU-Pro.}
    \label{tab:mmlu}
    \begin{tabular}{l||c|c|c|c|c|c}
        \hline
                      & \multicolumn{2}{c|}{\textbf{w/o}} & \multicolumn{2}{c|}{\textbf{prompt}} & \multicolumn{2}{c}{\textbf{FC}}                                \\
                      & score                             & time (m)                             & score                           & time (m) & score  & time (m) \\\hline
        Llama-3.1-8B  & 0.4440                            & 13.80                                & 0.2235                          & 25.45    & 0.1506 & 25.63    \\
        Llama-3.1-70B & 0.6231                            & 35.72                                & 0.5852                          & 64.50    & 0.5096 & 83.73    \\
        Hermes-3-8B   & 0.4122                            & 16.45                                & 0.4135                          & 18.80    & 0.3998 & 30.28    \\
        Mistral-22B   & 0.4993                            & 35.28                                & 0.3778                          & 54.65    & 0.4938 & 97.37    \\\hline
    \end{tabular}
\end{table*}

\subsection{Setup}

We use the same models and hyperparameters as in \S\ref{sec:setup}.
Likewise, we derive the same set of functions and system prompts to work as
two different robustness enhancement strategies, i.e., FC-based and
prompt-based enhancements. Besides, we also report the malicious detection
rate of the LLMs without any instruction (i.e., w/o) for comparison.

In this experiment, two datasets are used.
Wildjailbreak~\cite{wildteaming2024} is employed to systematically gauge
the effectiveness of different enhancement strategies since it contains a large
number of high-quality, diverse, and manually scrutinized malicious inputs.
In addition, we employ MMLU-Pro~\cite{wang2024mmlu}, a widely adopted
benchmark for evaluating the performance of LLMs. This dataset is used to
assess different enhancements' overheads in terms of inference time
and output quality (defined as the helpfulness score of the model). 

\subsection{Results}

\begin{table}[!tpb]
    \centering
    \small
    \caption{Effectiveness of different enhancements.}
    \label{tab:pilot}
    \begin{tabular}{l||c|c|c}
        \hline
                      & \textbf{w/o} & \textbf{prompt} & \textbf{FC} \\\hline
        Llama-3.1-8B  & 0.7424       & 0.8699          & 0.9943      \\
        Llama-3.1-70B & 0.4796       & 0.8133          & 0.9831      \\
        Hermes-3-8B   & 0.0531       & 0.1440          & 0.8492      \\
        Mistral-22B   & 0.0825       & 0.5915          & 0.6817      \\\hline
    \end{tabular}
\end{table}

\parh{Malicious Detection Capability.}~\T~\ref{tab:pilot} shows the
performance across different models after applying various enhancements.
From this table, it is evident that the FC-based enhancement is effective
in preventing the LLMs from generating malicious outputs across all four
models. 

Across all models, the FC-based enhancement exhibits superior performance
compared to the prompt-based strategy, with over 135\% improvement in the
malicious detection rate on average. Considering the semantic equivalence
between the FC-based and prompt-based enhancements, this result indicates
that FC can effectively enhance the LLMs' ability to comply with the
developers' instructions, thereby improving the LLMs' safety robustness.

\parh{Overhead.}~\T~\ref{tab:mmlu} shows the helpfulness of the model after
applying different enhancements. Llama family models exhibit a relatively
low tolerance to the additional enhancement. In contrast, the FC-based
enhancement exhibits better performance on other models, with a negligible
decrease (less than 0.02) in the helpfulness score compared to the
baseline. We presume that there are two main reasons for the differences.
First, the training preferences of models vary, which in turn affects the
helpfulness of the model. Compared with models with the same architecture
but different training data (Hermes-3-8B), it is evident that Llama family
models are inclined to be more sensitive to potential threats. Second, the
functions used in this experiment are merely a prototype designed to
demonstrate the feasibility of FC in enhancing LLM safety robustness. With
the development of more specific and delicate functions, we presume that
the helpfulness of the model can be further improved. 

The time overhead of different enhancements is also reported in
\T~\ref{tab:mmlu}. It is measured in minutes, reflecting the time required
for the model to address all the questions in the MMLU-Pro benchmark.
Despite the fact that the FC-based enhancement induces a general increase
in the time overhead, we note that the increase is acceptable. Given the
massive number of questions in the MMLU-Pro benchmark, the average
processing time for one question is approximately 0.5 seconds even for the
most time-consuming case (Mistral-22B with FC). We attribute the growth in
time overhead to the increase in the number of tokens in the system prompt,
as the specifications of the functions are appended to it. Likewise, we
believe that future work can smoothly condense the function specifications
to optimize over our prototype.

To conclude, this experiment demonstrates the promising safety robustness
enhancement capability of FC in LLMs, further validating our findings from
the causality analysis\textemdash FC can effectively enhance the LLMs'
ability to comply with instructions.

\section{Related Work}
\label{sec:related}

Causality analysis is a canonical approach to investigate the internal
mechanism of neural networks, as a series of research has demonstrated its
effectiveness. Chattopadhyay et al.~\cite{chattopadhyay2019neural} first
proposed the equivalence between the neural network and the SCM, laying the
foundation for the subsequent causality analysis of neural
networks~\cite{vig2020investigating, sun2022causality, ji2023cc}. Besides,
causality is also widely employed to guide the model edits to improve LLM's
performance~\cite{meng2022locating, meng2023memit, fang2024alphaedit,
li2024badedit}, demonstrating the practical application of causality in the
field of LLMs.


\section{Conclusion}

In this paper, we presented a causality-based investigation of function
calling in large language models, conducting layer- and token-level
interventions to uncover its substantial influence on internal
computational logic. We demonstrated, through benchmark experiments on
safety robustness enhancement across four mainstream LLMs, that function
calling yields an average performance improvement of 135\% over
conventional prompting in detecting adversarial inputs. These findings
reveal the mechanistic underpinnings of function calling and underscore its
potential for enhancing LLM reliability and capability in real-world
applications.

\bibliography{ref}


\end{document}